\documentstyle[prl,aps]{revtex}

\input epsf
\begin{document}
\draft \twocolumn[\hsize\textwidth\columnwidth\hsize\csname
@twocolumnfalse\endcsname

\title{Convective instability on a crystal surface}
\author{Navot Israeli and Daniel Kandel}
\address{Department of Physics of Complex Systems, Weizmann Institute
of Science, Rehovot 76100, Israel}
\author{Michael F. Schatz and Andrew Zangwill}
\address{School of Physics, Georgia Institute of Technology,
Atlanta, GA 30332} 
\maketitle


\begin{abstract}

The distinction between absolute and convective instabilities is well
known in the context of hydrodynamics and plasma physics. In this
Letter, we examine an epitaxial crystal growth model from this point of
view and show that a strain-induced step bunching instability can be
convective. Using analytic stability theory and numerical simulations,
we study the response of the crystal surface to an inhomogeneous
deposition flux that launches impulsive and time-periodic perturbations
to a uniform array of steps. The results suggest a new approach to
morphological patterning.

\end{abstract}

 ]


Instabilities dominate the behavior of many nonlinear systems.
In hydrodynamics \cite{Godreche} and plasma physics \cite{Bers,L&L}, it
is commonplace to distinguish
between an {\it absolute} and a {\it convective} instability. In the
first case,
an initial perturbation spreads more rapidly than it advects and the
system
evolution is insensitive to subsequent perturbations. In the second
case, an initial
perturbation advects more rapidly than it spreads and the system
evolution
remains sensitive to subsequent perturbations. The latter characteristic
presents
an opportunity for control. In this Letter, we show that a
strain-induced instability that can occur during epitaxial growth
is convective and suggest a novel approach
to morphological patterning based on the implied control.

We consider a regular, staircase-like surface composed of flat
terraces of average width $\ell$. The terraces are separated by
straight, parallel, atomic height steps with horizontal positions $x_n$,
where the index $n$ grows in the direction of negative surface slope.
We assume that atoms impinge on each lattice site at a rate $F$. This
leads to a build up of a finite concentration of adatoms on the
terraces. Adatoms diffuse on terraces and attach to the bottom of steps
at a rate $K$. Atoms can also detach from steps towards neighboring
terraces. These processes lead to step motion which can be described by
simple equations \cite{BCF}. If the flux is large enough, the steps
acquire a net
positive velocity, inducing vertical growth of the crystal by one atomic
unit after every step has moved a distance of one terrace width. 

The equations of motion for the steps are much more complicated if
there are long-range interactions between
steps. An example is the growth of a strained film where
the lattice constant of the deposited material differs from the
lattice constant of the substrate. The corresponding equations of
motion \cite{Tersoff} can be simplified if we assume that diffusion is
fast.
In this limit, the step velocities are given by
\begin{equation}
v_n=\frac{F}{2}\left(x_{n+1}-x_{n-1}\right)+\frac{K}{2}\left(\mu_{n+1}+
\mu_{n-1}-2\mu_n\right)~,
\label{step_velocity}
\end{equation}
where $\mu_n$ is the chemical potential associated with adding an atom
to the solid at the $n$th step. The $\mu_n$ are
\begin{equation}
\mu_n=\sum_{m\neq
n}\left(\frac{\beta}{\left(x_m-x_n\right)^3}-\frac{\alpha}{x_m-x_n}
\right)\;,
\label{chemical_potential}
\end{equation}
where $\alpha$  reflects the attractive interaction arising from the
elastic relaxation around each step on a strained layer, and $\beta$
reflects the repulsion arising from the inherent stress of each step.

One solution to this model is uniform step-flow, where every
step moves with velocity $F\ell$. Under certain circumstances, this
steady state becomes unstable and groups of steps bunch together
\cite{Tersoff}.
In this paper, we analyze the bunching instability from a new point of
view.

The distinction between an absolute and a convective
instability is most significant for a problem with at least one
preferred frame of reference. For our problem, the {\it lab frame}
is one such frame. We will also be interested in perturbing the step
train by supplementing the uniform deposition flux with a very
narrow beam of atoms that can be moved across the surface. This
beam is at rest in the {\it source frame} of reference.

\begin{figure}[h]
\centerline{ \epsfxsize=80mm \epsffile{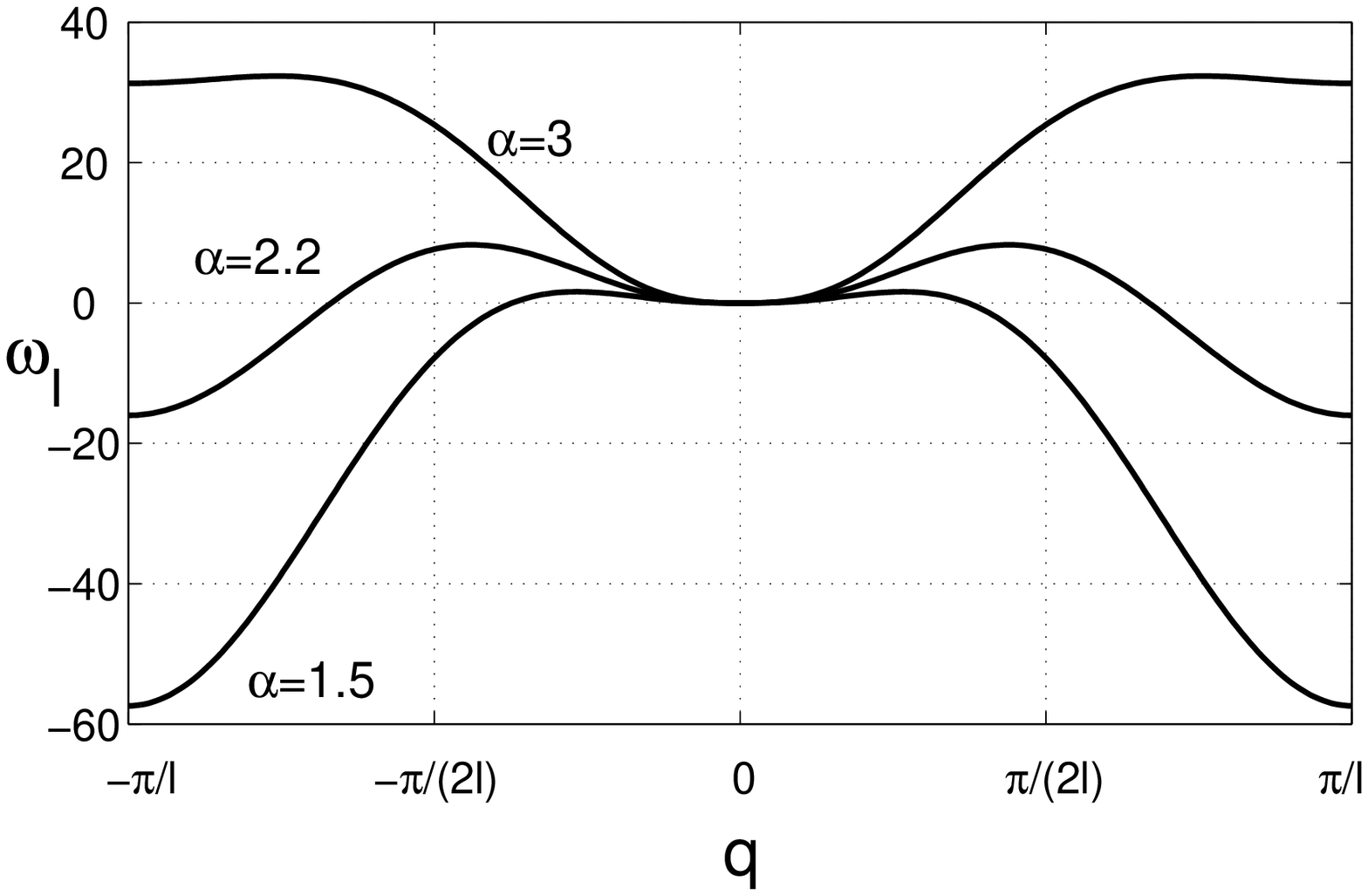}}
\caption{$\omega_I(q)$ for real $q$ and different values of
$\alpha$. The values of the other parameters in units where $\ell=1$ are
$F=10$, $K=6$, $\beta=1$ and $M=299$.} \label{figure1}
\end{figure}

In the lab frame, the linear stability of uniform step-flow motion
against a
perturbation of the step positions $\delta x_n(t)=\epsilon \exp
\left[i \left(n\ell q-\omega t\right)\right]$ leads to the dispersion
relation
\begin{eqnarray}
\label{dispersion} &D&_{lab}\left(q,\omega\right) = -i\left(\omega-F\ell
q+ F \sin{\ell q}\right)  \\ &-& 2K\left(\cos{\ell q}
-1\right)\sum_{m=1}^M\left(\cos{m\ell q}-1\right)\left(\frac{\alpha}
{m^2 \ell^2}-\frac{3\beta}{m^4\ell^4}\right)\;. \nonumber
\end{eqnarray}
Here $M$ is the number of neighbors a step
interacts with on each side. In a general frame of reference,
$D(q,\omega)=D_{lab}(q,\omega+qv_f)$, where $v_f$ is the velocity
of the frame of reference with respect to the lab frame.
Conventional stability theory seeks the complex zeroes $\omega(q)=
\omega_R(q) + i
\omega_I(q)$ of
$D(q,\omega)$ for given real $q$. $\omega_I(q) > 0$ is a sufficient
condition for instability of uniform step-flow. Figure
\ref{figure1} shows $\omega_I(q)$ of our model for different
values of $\alpha$. The $q=0$ mode is marginal for all values of
$\alpha$. When $\alpha>0$ is small there are two additional
marginal modes with $q=\pm q_m$. All the modes with $-q_m<q<q_m$
(except for $q=0$) are unstable. When $\alpha$ is increased, $q_m$
increases towards $\pi/\ell$ and the interval of unstable modes becomes
wider.

To distinguish between different types of instability,  we
consider the long-time behavior of the mode with wave-number $q^0$
that has a zero group velocity: ${\partial \omega_R(q) / \partial
q}|_{q=q^0} = 0$. By definition, the system is absolutely
unstable if $\omega_I\left(q^0\right)
> 0$ and convectively unstable if $\omega_I\left(q^0\right) < 0$.  In
physical terms, this is equivalent to examining the
time-asymptotic behavior of a disturbance launched by an
impulse-type (localized in space and time) perturbation. When the
instability is absolute, the disturbance spreads in space more
rapidly than it advects; an observer at any fixed position sees
asymptotic growth since $\omega_I\left(q^0\right) > 0$. When the
instability is convective, the disturbance advects more rapidly
than it spreads; an observer at any fixed position may see
transient growth as the disturbance advects by, but finds
asymptotic decay since $\omega_I\left(q^0\right)< 0$.

\begin{figure}[h]
\centerline{ \epsfxsize=80mm \epsffile{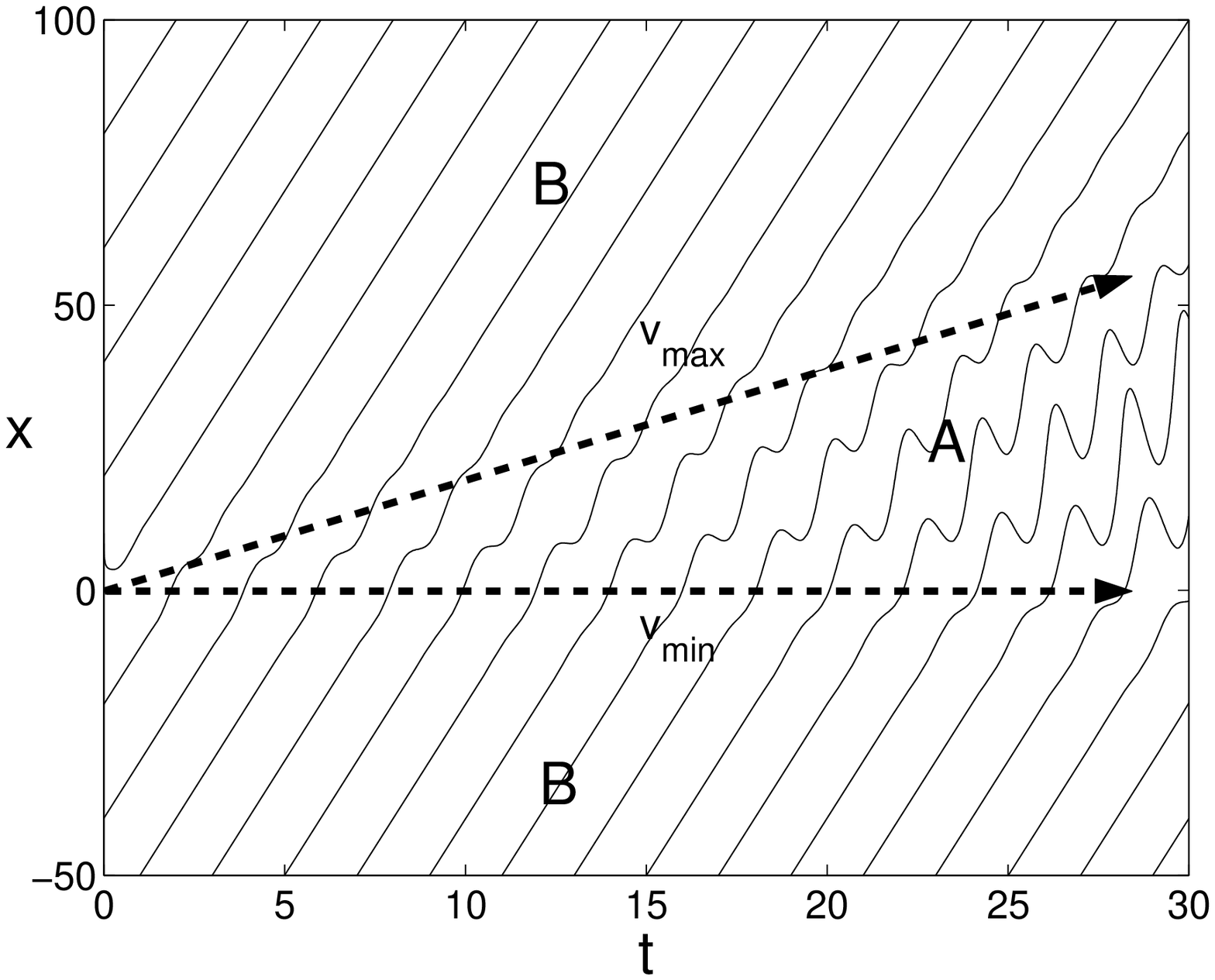}}
\caption{Space time plot of a system of 300 steps with periodic
boundary conditions after the application  of an impulsive
perturbation to a single step at $x=0$ and $t=0$. Each line shows
the trajectory of a single step in the lab frame. Only a small
portion of the system is plotted and the step motion is amplified.
The choice of parameters for this specific system in units where
$\ell=1$ is $F=10$,
$K=6$, $\alpha=0.9$, $\beta=1$ and $M=10$. } \label{figure2}
\end{figure}

Figure \ref{figure2} is a space-time plot of step trajectories (in
the lab frame) determined by a numerical solution of the equations
of motion with an impulsive perturbation applied at $t=0$ to a
single step. In the laboratory, a perturbation of this kind can be
generated using the narrow beam source mentioned above
\cite{caveat}. The perturbation has no effect on uniform step flow
in the portion of the surface labeled Region B in Fig.\
\ref{figure2}. However, in Region A, the perturbation creates a
disturbance that spreads and amplifies in the direction of step
flow. $v_{min}$ and $v_{max}$ are the minimal and maximal group
velocities (in the lab frame) of unstable Fourier modes (for which
$\omega_I(q)\geq0$). As it happens, $v_{min}=0$ for this model so the
disturbance neither spreads out over the entire crystal nor
advects away from the point where the impulse was applied. In
other words, step bunching as observed in the lab frame is ``on
the border'' between absolute and convective instability. This
fact can be used to test the step-bunching model experimentally
because it does not depend on any of the model parameters. By
contrast, the transition between absolute and convective
instability in hydrodynamic systems typically occurs at a single
value of the control parameter.

\begin{figure}[h]
\centerline{ \epsfxsize=80mm
\epsffile{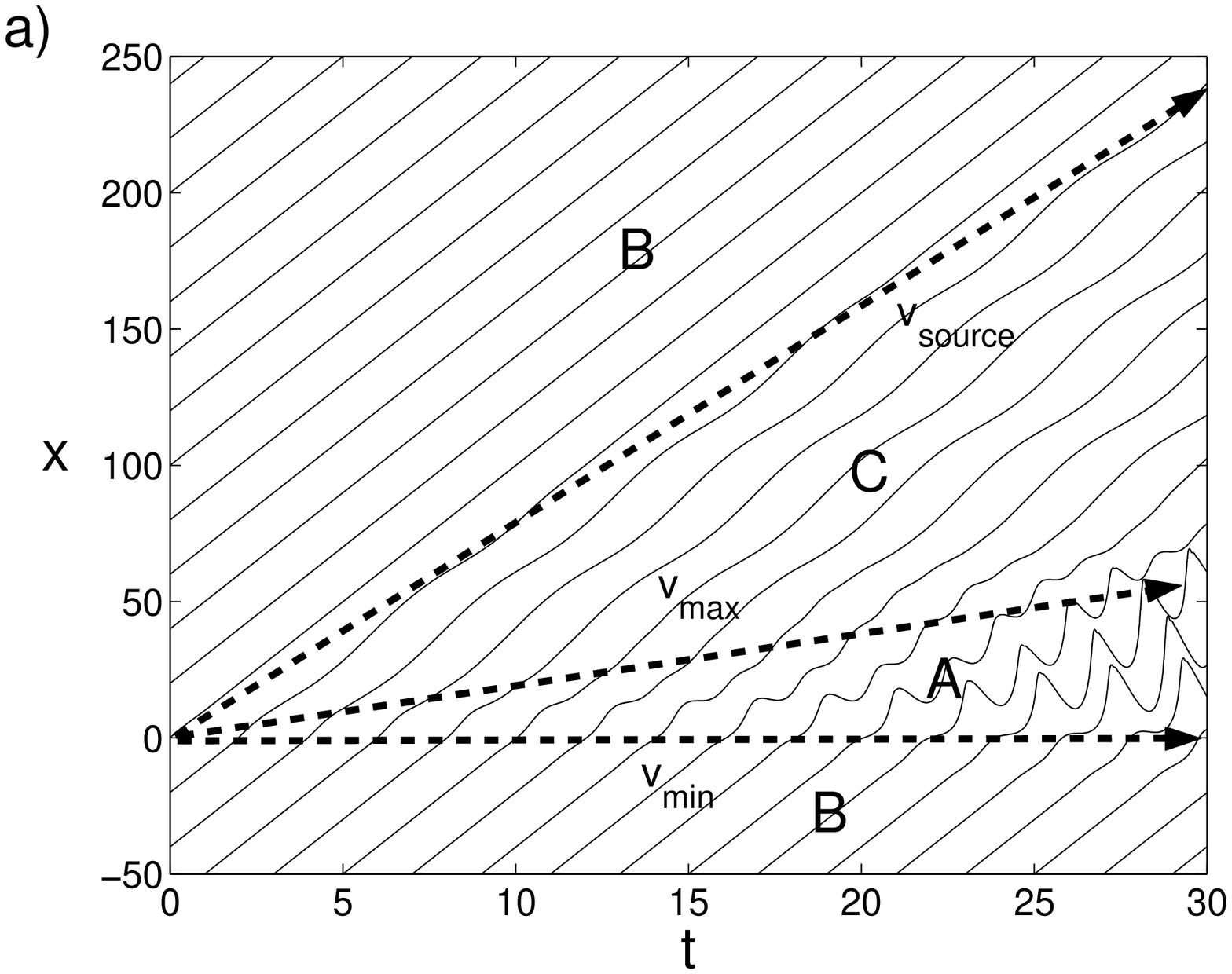}}
\centerline{\epsfxsize=80mm
\epsffile{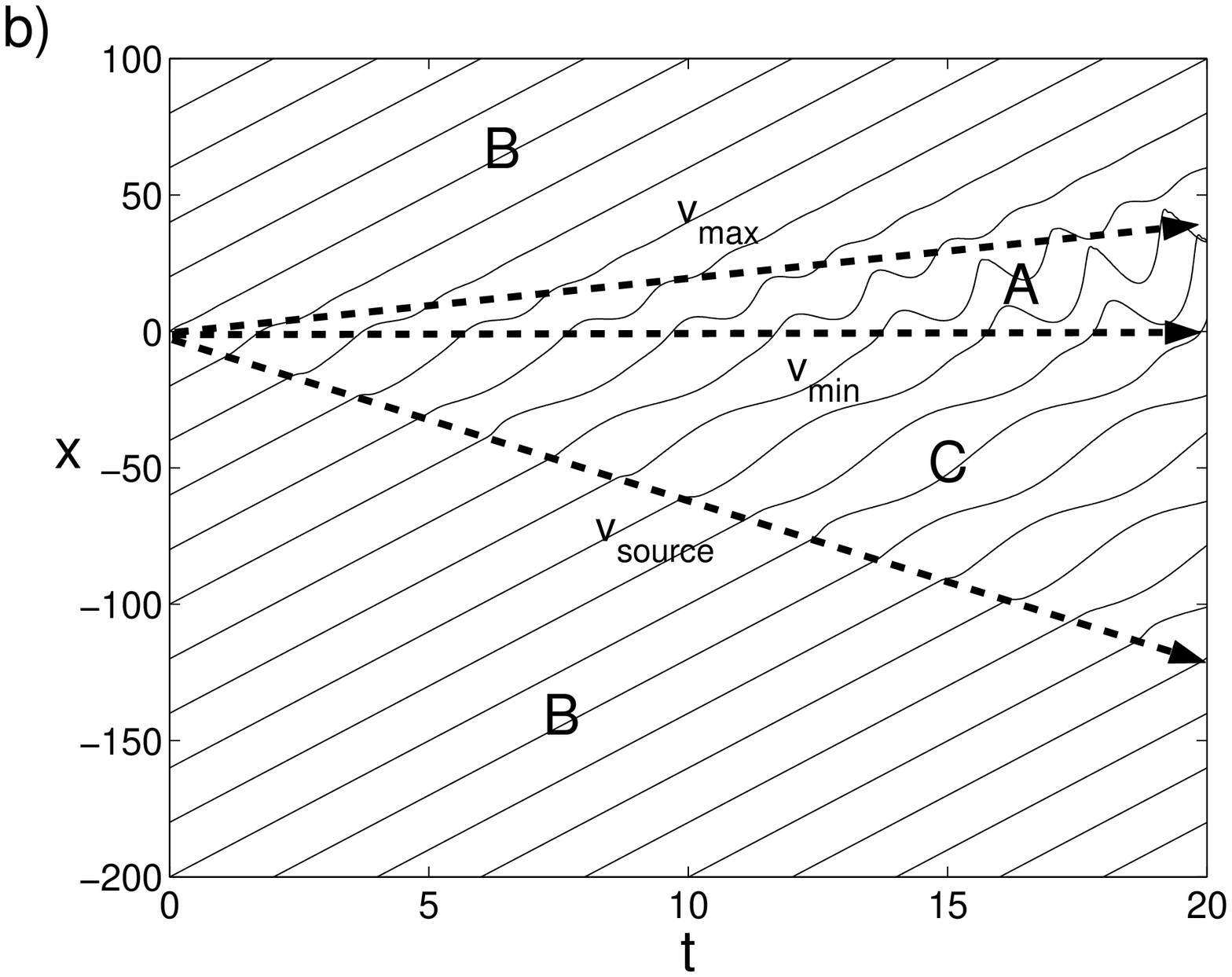}}
 \caption{Space time plots of systems of 300
steps with periodic boundaries perturbed by a narrow harmonic
source. Each line shows the trajectory of a single step in the lab
frame. Only a small portion of the system is plotted and the step
motion is amplified. We have indicated the rays which correspond
to the source velocity and the velocities $v_{min}$ and $v_{max}$. When
the source velocity $v_{source}>v_{max}$ and
$|\omega_s|<\omega_c$, periodic step bunching is amplified in Region C
in the
direction opposite to step flow (a). When $v_{source}<v_{min}$ and
$|\omega_s|<\omega_c$, periodic step bunching is amplified in Region C
along
the direction of step flow (b). The choice of parameters for these
specific systems in units where $\ell=1$ is $F=10$, $K=6$, $\alpha=0.9$
and $\beta=1$. The source width is $a=\ell$.} \label{figure3}
\end{figure}

We turn next to the response of the growing crystal to spatially
localized but {\it time-periodic} perturbations produced by the narrow
beam source. Such perturbations generate two types of asymptotic
behavior which we will call {\it switch-on bunching} and {\it
time-periodic bunching}. If the source moves with velocity $v_s$ (in the
lab frame) in the interval $v_{min}<v_s<v_{max}$, the system is
absolutely unstable in the source frame and switch-on bunching occurs
exclusively. For this situation, the step pattern develops analogously
to the impulsive case (Fig. \ref{figure2}). However, if $v_s>v_{max}$ or
$v_s<v_{min}$, the system is {\em convectively} unstable in the source
frame. Switch-on bunching still occurs on one portion of the crystal
surface, but, in addition, time-periodic bunching (which is sensitive to
the nature of the forcing) {\it can} occur on an adjacent portion of the
surface (Region C in Fig.\ \ref{figure3}).

To determine whether time-periodic bunching {\it does} occur in the
regime of convective instability, it is sufficient to examine the
asymptotic linear response of the step system to a spatially
localized, time-harmonic source,
\begin{equation}
S_n\left(t\right)=\exp \left(-\frac{4\left[n\ell-(v_s-F\ell)
t\right]^2}{a^2}-i\omega_s t\right)~, \label{source}
\end{equation}
where $\omega_s$, $a$ and $v_s$ are the source frequency, width and
velocity in the lab frame. This is called the {\it signaling
problem} \cite{Godreche,L&L}. In the source frame ($n_s =
n-(v_s-F\ell)t/\ell$) we find that
\begin{equation}
\delta x_{n_s}(t)\propto \exp (-i\omega_s t)\int_C \frac{\exp
(in_s\ell q-a^2 q^2)}{D_{lab}(q,\omega_s+v_sq)}dq\;,
\label{integral_xnt}
\end{equation}
where $D_{lab}$ is the dispersion relation (\ref{dispersion})
continued to the complex $q$ plane and $C$ is a suitable contour
in this plane. The zeroes of $D_{lab}(q,\omega_s+v_sq)$ in the $q$ plane
dominate the integral. Among these, the most important zero corresponds
to the single mode whose wave-vector $q^*(\omega_s)$ has a real part
$q^*_R(\omega_s)$ in the interval
$[-\pi/\ell,\pi/\ell]$ and an imaginary part
$q^*_I(\omega_s)$ that can change sign as $\omega_s$ changes.

The main result is that there exists a critical frequency
\begin{equation}
\omega_c=\left|F\sin \ell q_m +q_m \left(v_s-F\ell\right)\right|\;.
\end{equation}
If $|\omega_s|>\omega_c$, the amplitude of time periodic step
bunching decays as the distance from the source increases. The
source has little effect on the step-flow pattern in this case. However,
if
$|\omega_s|<\omega_c$, the source induces time-periodic step
bunching that grows exponentially in space:
\begin{equation}
\label{nt_response2} \delta x_{n_s}(t) \propto \left.
\frac{\exp\left[in_s\ell q-i\omega_st-a^2
q^2\right]}{\frac{dD_{lab}(q,\omega_s+v_sq)}{dq}}
\right|_{q=q^*(\omega_s)}\;.
\end{equation}
There are two cases to consider. If $v_s>v_{max}$, the disturbance
grows in the direction opposite to step flow because
$q^*_I(\omega_s)>0$ (Region C of Fig.\
\ref{figure3}(a)). Conversely, if $v_s<v_{min}$, we find
$q^*_I(\omega_s)<0$ and the disturbance grows along the direction
of step flow (Region C in Fig.\ \ref{figure3}(b)). Regions A and
B correspond to switch-on bunching and uniform step flow similar
to the corresponding regions in Fig.\ \ref{figure2}.

For a step-bunch that grows from a time-harmonic perturbation,
there is very little nonlinear distortion of the bunch shape close
to the source, as expected. Frequency spectra collected at
different spatial locations in Region C show that higher harmonics
contribute more as the distance from the source increases.
Nevertheless, the amplitudes of the fundamental and all higher
harmonics {\it saturate} for distances sufficiently far from the
source. 

The stabilizing influence of nonlinearity
in the step-flow case prevents the system from wandering too far
away from the linear character of the imposed perturbation. This
provides an opportunity to exploit the bunching instability to
intentionally pattern the crystal surface in Region C.
The idea is to apply a perturbation prepared as a superposition of
terms of the form (\ref{source}) with frequencies in the range
$|\omega_s|<\omega_c$. As long as nonlinear effects can be
ignored, each of these terms evolves according to Eq.\
(\ref{nt_response2}) in Region C. We
can therefore tune the values of the amplitudes and phases of
the various terms of the perturbation, so that at a specific time the
step
configuration in Region C would be close to a pre-designed
morphology.

\begin{figure}[h]
\centerline{ \epsfxsize=80mm \epsffile{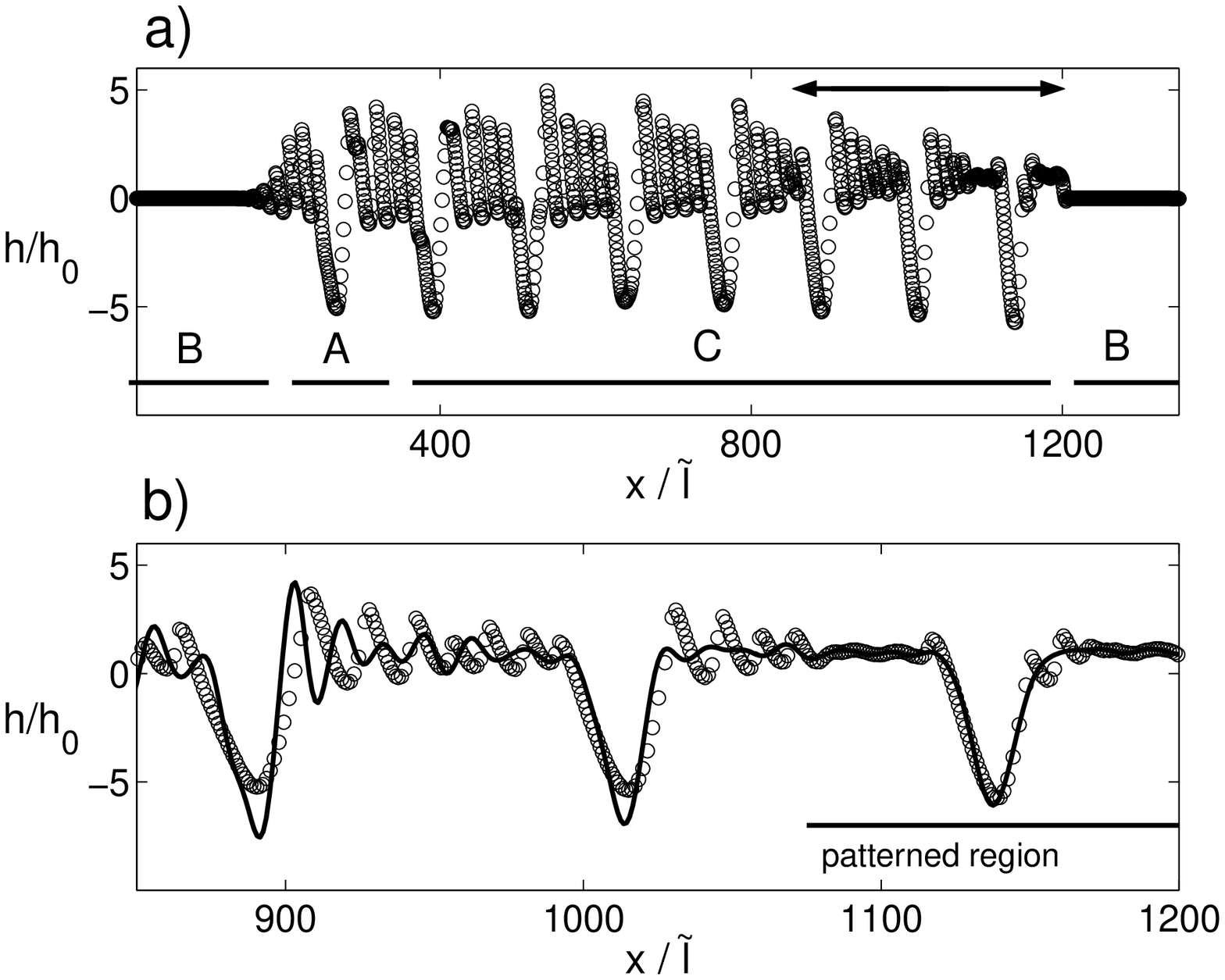}}
\caption{Surface height (in units of the height of a single step, $h_0$)
as a function of position evolving from a superposition of sinusoidal
disturbances. $h=0$
corresponds to the plane $x_n=n\ell$ and the unit of length is the
average step seperation on this plane,
$\tilde{\ell}=\sqrt{\ell^2+h_0^2}$. The source velocity
is $v_s=F\ell$ and it is located at $x/\tilde{\ell}=1200$. Regions A, B
and C in (a) are
analogous to the corresponding regions in Fig.\ \ref{figure3}(a). (b) is
a magnification of
the region marked by the two-sided arrow in (a).
The solid line in (b) shows the predicted linear response, while the
actual
surface morphology, resulting from the solutions of the step equations
of motion, is shown as circles.} \label{figure4}
\end{figure}

In order to demonstrate the applicability of this idea, we attempted to
induce a groove-like pattern in a region which contained 125 steps. To
construct this
pattern, we used the linear analysis to optimize a small set of
amplitudes and phases for
waves with frequencies in the range $|\omega_s|<\omega_c$. We then
numerically solved the step equations of motion with the
designed source. The resulting surface height as a function of position
is shown as circles in Fig.\ \ref{figure4}. In Fig.\ \ref{figure4}(a) we
have indicated regions analogous to Regions A, B and C of Fig.\
\ref{figure3}(a). Figure \ref{figure4}(b) is a magnification of the
section marked by a two-sided arrow in Fig.\ \ref{figure4}(a). The
region of 125 steps we attempted to pattern is marked. The
solid line in Fig.\ \ref{figure4}(b) shows the predicted linear response
of the surface to the designed source. Inside the patterned region it is
very similar to the desired pattern. Near the source
($x/\tilde{\ell}=1200$),
the shape of the surface obtained from the numerical
solution of the step equations of motion closely follows the linear
dynamics. Further from the source, nonlinearity acts and we
observe a regular sequence of grooves which
are recognizably ``echoes'' of the original pattern for many periods. We
have checked that this
behavior is robust in the presence of deposition shot noise.

In a typical experimental system the dispersion relation is not known.
Nevertheless, one can in principle investigate the response
of the step system to sources of different frequencies experimentally.
For each
frequency, one can measure the induced wavelength ($q^*_R$) and
amplification rate ($q^*_I$), as well as the prefactor multiplying the
exponential in Eq.\ (\ref{nt_response2}).
This information is sufficient for the implementation of the design
procedure outlined above.

In summary, we have demonstrated the convective nature of a
step-bunching instability
that occurs in a recently proposed model of epitaxial, strain-induced,
step-flow growth.
A variety of step bunching scenarios arise when conventional
step-flow is perturbed by a beam of atoms whose flux can be controlled
as a function
of space and time. In particular, there is a regime of time-periodic
bunching
that can be used to launch a sequence of pre-designed step-bunch
patterns.
The nonlinearity of the model is such that the bunches do not distort
appreciably as growth proceeds.

D.K. acknowledges the support of the Israel Science Foundation.
M.F.S. is a Cottrell Scholar of the Research Corporation and gratefully
acknowledges the support of the National Science Foundation under Grant
CTS-9876590.  A.Z. gratefully acknowledges the support of the Department
of Energy under Contract DE-FG02-97ER45658.


\begin{thebibliography}{99}
\bibitem{caveat} Our main conclusions should remain valid even if
many steps are perturbed in an actual experiment.
\end{thebibliography}
\end{document}